\documentclass[24pt]{article}
\usepackage{amssymb}
\usepackage{amsfonts}
\usepackage{amssymb,amsmath}
\usepackage{mathrsfs}
\usepackage{latexsym}
\usepackage{amsmath}
\usepackage{latexsym}
\usepackage[cp1251]{inputenc}
\usepackage{graphicx}
\usepackage{color}

\title{An application of a quantum wave impedance approach  for solving a nonsymmetric single well problem}
\author{O. I. Hryhorchak\\
{\small Department for Theoretical Physics, Ivan Franko National
University of Lviv,}\\
{\small 12, Drahomanov Str., Lviv, UA--79005,
Ukraine}\\
\small{\it{orest.hryhorchak@lnu.edu.ua}}}

\def\tg{\mathop{\rm tg}\nolimits}

\begin{document}
\renewcommand{\abstractname}{Abstract}
\maketitle

\begin{abstract}
A short introduction of a relation between  a Green's function and a quantum wave impedance function as well as its application to a determination of eigenenergies and eigenfunctions of a quatum-mechanical system is provided. Three different approaches, namely a classical approach based on a direct solving of a Shr\"{o}dinger equation, a transfer matrix method and a quantum wave impedance technique, for a calculation of eigenenergies and eigenfunctions of a quantum mechanical nonsymmetric single well system are considered. A comparision of these approaches gives the possibility to clarify advantages and drawbacks of each method which is useful especially for teaching and learning purposes.
\end{abstract}

\section{Introduction}
For the first time an introduction of a quantum wave impedance concept  was done in 1988  by the authors of the pioneer article on this topic \cite{Khondker_Khan_Anwar:1988}. Accordingly to their statement the primary purpose of this article was to show
that the well-developed theory of electrical transmission lines can be
effectively used for calculating the quantum mechanical transmission probability.  One year later the same authors have published an article \cite{Anwar_Khondker_Khan:1989} in which they calculated the traversal time of electrons in resonant
tunnelling structures and showed that the real part of the
quantum-mechanical wave impedance, at resonance,
can be used to calculating the electron traversal time.
They concluded and emphasized the usefulness of a quantum-mechanical impedance concept and the simplicity of using the transmission line equation.

In many other papers \cite{Khondker:1990, Morrisey_Alamb_Khondker:1992, Khondker_Alam:1991,  Khondker_Alam:1992, Kaji_Hayata_Koshiba:1992, Nelin:2012, Nazarko_etall:2015, Ashby:2016, Nelin_Zinher_Popsui:2017,  Nelin_Shulha_Zinher:2017}  the efficacy of a quantum wave impedance approach for an analysis of quantum-mechanical structures with a potential which has a complicated spatial structure was demonstrated but only in the paper \cite{Arx1:2020} it was shown how on the base of a the Shr\"{o}dinger equation to get an equation for a quantum wave impedance function and to reformulate the scattering and bound states problems in terms of a quantum wave impedance. 
  
   Despite the fact just mentioned the relation between a Green's function and a quantum wave impedance function was established more quite long ago \cite{Khondker_Alam:1991}. The approach of Green's functions is very useful for solving non-homo\-geneous boundary value problems which is very close, from a mathe\-matical point of view, to the main reason of a quantum wave impedance introduction.

A few years later in  \cite{Hague_Khondker:1998}  a technique of a calculation of the normalized wave functions in arbitrary one-dimen\-sional quantum well structures was described. Authors using the relation between Green's function and a quantum wave impedance presented an efficient technique of calculating both
the eigenenergies and the normalized
eigenfunctions in quantum wells. This method
is particularly attractive in numerical calculations of multibarrier devices in which the estimation of the self-consistent potential is desired. The method is computationally efficient and is generalized enough to model arbitrary wells under an applied bias voltage including the effects of space charge.
 
The aim of this article is to provide the comparison of a quantum wave impedance approach with a classical one (based on a direct solving of a Shr\"{o}dinger equation) and with a transfer matrix method. We will apply mentioned approaches specifically for finding energies and wave functions of bound states. It will help one to clearly understand advantages and drawbacks of a quantum wave impedance method.

To fulfil our task we will consider very simple system. But it is enough to understand the main features of each approach. So, assume that we have a nonsymmetric rectangular potential well of a finite depth: 
\begin{eqnarray}\label{UU1U2U3}
U(x)=\left\{\begin{array}{cc}
U_1>0, & x\leq 0\\
U_2<0, &  0<x<a\\
U_3>0, & x>a
\end{array}\right..
\end{eqnarray} 
We are going to find energies and wave functions of bound states and these states have an energy $E<max(U_1, U_3)$. 
 
 \section{The relation between a Green's function and a quantum wave impe\-dance}

 The equation for a Green's function $G(x,x',E)$ is as follows
 \begin{eqnarray}\label{Green_equation}
 \left(E+\frac{\hbar^2}{2m}\frac{d^2}{dx^2}-U(x)\right)G(x,x',E)=\delta(x-x').
 \end{eqnarray}
 It coincides with the Shr\"{o}dinger equation everywhere besides point $x=x'$ and this fact pushes us to introduce 
 an extended quantum wave impedance function $\tilde{Z}(x,x')$ in the following way 
 \begin{eqnarray}
 \tilde{Z}(x,x')=\frac{\hbar}{im}\frac{\partial G(x,x',E) / \partial x}{G(x,x',E)}.
 \end{eqnarray}
 Then on the base of the equation (\ref{Green_equation}) we get 
 \begin{eqnarray}
 \frac{d\tilde{Z}(x,x')}{dx}+i\frac{m}{\hbar}\tilde{Z}^2(x,x')=i\frac{2}{\hbar}(E-U(x))-i\frac{2}{\hbar}\frac{\delta(x-x')}{G(x,x',E)}.
 \end{eqnarray}
 If $U(x)$ is not singular then after integration of both left and right sides of the previous equation we obtain:
 \begin{eqnarray}
 Z^+(x')-Z^-(x')=i\frac{2}{\hbar}G^{-1}(x',x',E)
 \end{eqnarray}
 or
 \begin{eqnarray}
 G(x',x',E)=-i\frac{2}{\hbar}\frac{1}{Z^+(x')-Z^-(x')},
 \end{eqnarray}
 where
 \begin{eqnarray}
 Z^-(x')=\tilde{Z}(x'+0,x'),\qquad
 Z^+(x')=\tilde{Z}(x'-0,x').
 \end{eqnarray}
 If we apply boundary conditions for $\tilde{Z}(x,x')$ at points with coordinates $a$ and $b$ then instead of $Z^-(x')$ and $Z^+(x')$ we can use $Z^-(x',b)$ and $Z^+(x',a)$, where $a<x'<b$. Notice that $Z^-(x',b)$ and $Z^+(x',a)$ are the same ones as in a \cite{Arx1:2020}. Now we can transit the well-know relations obtained within Green's function approach to a formalism of a quantum wave impedance.
  
 Now let's consider the following relation: 
 \begin{eqnarray}
 G(x,x',E+i\varepsilon)=\sum_m\frac{\psi_m(x)\psi_m^*(x')}{E-E_m+i\varepsilon},
 \end{eqnarray}
 where $\varepsilon\rightarrow0$ and thus
 \begin{eqnarray}
 \lim_{E\rightarrow E_n} G(x,x,E+i\varepsilon)=\frac{|\psi_n(x)|^2}{E-E_n+i\varepsilon}.
 \end{eqnarray}
 Taking an imaginary part of both sides of the previous relation we get
 \begin{eqnarray}
 |\psi_n(x)|^2=-\varepsilon\Im\left[G(x,x,E_n-i\varepsilon)\right]
 \end{eqnarray}
 or using a quantum wave impedance we obtain
 \begin{eqnarray}\label{psiZ+Z-}
 |\psi_n(x)|^2=\frac{2\varepsilon}{\hbar}\Im\left(\frac{i}{Z^-(x,b,E_n-i\varepsilon)-Z^+(x,a,E_n-i\varepsilon)}\right).
 \end{eqnarray}

\section{Nonsymmetric well of a finite depth. Quan\-tum wave impedance approach}
Characteristic impedances for each region of a potential (\ref{UU1U2U3}) are $z_1=\frac{\hbar \varkappa_1}{m}$, $z_2=\frac{i\hbar k_2}{m}$, $z_3=\frac{\hbar \varkappa_3}{m}$. The load impedance coincides with the characteristic impedance of a third region $z_3$. Calculating the input impedance at a point $x=0$, on the base of a well-known formula \cite{Khondker_Khan_Anwar:1988}, and equating it to $-z_1$ \cite{Arx1:2020} we easily get the expression for determining energies of bound states:
\begin{eqnarray}
-z_1=z_2\frac{z_3\cos[k_2a]-z_2\sin[k_2a]}
{z_2\cos[k_2a]-z_3\sin[k_2a]}=z_2\frac{z_3-z_2\tan[k_2a]}
{z_2-z_3\tan[k_2a]}
\end{eqnarray}
or reminding the relations between $z_m$ and $k_m$, $m=1,2,3$ we obtain the same formula as in two previous sections
\begin{eqnarray}
\tan(k_2a)=\frac{k_2(\varkappa_1+\varkappa_3)}{\varkappa_1\varkappa_3-k_2^2}.
\end{eqnarray}
It is easy to notice that a quantum wave impedance approach for the calculation of eigenenergies demands much less efforts than the classical method or a transfer matrix formalism.

To find wave functions of bound states in this system let's find the values of $Z^+(x,a;E)$ and $Z^-(x,b;E)$ in the second region of a potential ($\ref{UU1U2U3}$). It is easy to do \cite{Arx1:2020} since that region has a constant value of a potential energy. Thus,
\begin{eqnarray}
Z^-(x,a;E)=z_2\tan(ik_2x+\phi_R),\nonumber\\
Z^+(x,0;E)=z_2\tan(ik_2x+\phi_L),
\end{eqnarray}
where
\begin{eqnarray}
\!\!\!\phi_R\!\!\!&=&\!\!\!-\frac{1}{2}\ln\left(\exp(2ik_2a)\frac{z_2-z_3}{z_2+z_3}\right)=-\frac{1}{2}\ln\left(\exp(2ik_2a)\frac{k_2-ik_3}{k_2+ik_3}
\right),\nonumber\\
\!\!\!\phi_L\!\!\!&=&\!\!\!-\frac{1}{2}\ln\left(\frac{z_1+z_2}{z_2-z_1}\right)=-\frac{1}{2}\ln\left(\frac{ik_1+k_2}{k_2-ik_1}\right).
\end{eqnarray}
The Taylor series of $Z^+(x,0;E_n-i\varepsilon)$ and $Z^-(x,a;E_n-i\varepsilon)$ have the following form
\begin{eqnarray}
Z^-(x,a;E_n-i\varepsilon)&=&Z^+(x,E_n)+\frac{i}{\hbar}
\frac{\varkappa_3a+1}{\varkappa_3}\varepsilon+O(\varepsilon^2),\nonumber\\
Z^+(x,0;E_n-i\varepsilon)&=&Z^-(x,E_n)+\frac{i}{\hbar}
\frac{1}{\varkappa_1}\varepsilon+O(\varepsilon^2),
\end{eqnarray}
where $E_n$ we find from the condition $\phi_L=\phi_R$ and it gives the same relation (\ref{nsw_dr}) as we got in the previous sections. Now using the obtained earlier relation (\ref{psiZ+Z-}) we finally get
\begin{eqnarray}
|\psi_2(x)|^2=2\left(a+\frac{\varkappa_1+\varkappa_3}{\varkappa_1\varkappa_3}\right)^{-1}\cos^2(k_2a+\phi),
\end{eqnarray}
where $\phi\equiv \phi_L=\phi_R$.

 \section{Classical approach}
 Having the potential energy in a form (\ref{UU1U2U3}) we get the following Shr\"{o}dinger equations in each region
 \begin{eqnarray}
 \qquad -\frac{\hbar^2}{2m}\frac{\partial^2}{\partial x^2}\psi_1(x)+U_1\psi_1(x)&=&E\psi_1(x),\quad  x\leq0,\nonumber\\
 -\frac{\hbar^2}{2m}\frac{\partial^2}{\partial x^2}\psi_2(x)+U_2\psi_1(x)&=&E\psi_2(x), \quad 0<x\leq a,\nonumber\\
 -\frac{\hbar^2}{2m}\frac{\partial^2}{\partial x^2}\psi_1(x)+U_3\psi_1(x)&=&E\psi_3(x), \quad  x>a.
 \end{eqnarray}
 And general solutions of these equations (in each region) for an energy $E<max(U_1, U_3)$ are easy enough
 \begin{eqnarray}
 \psi_1(x)&=&C_{11}\exp(\varkappa_1x)+C_{12}\exp(-\varkappa_1x),\nonumber\\
 \qquad \psi_2(x)&=&C_{21}\cos(k_2x)+C_{22}\sin(k_2x),\nonumber\\
 \qquad \psi_3(x)&=&C_{31}\exp(\varkappa_3x)+C_{32}\exp(-\varkappa_3x),
 \end{eqnarray}
 where 
 \begin{eqnarray}
 \varkappa_1\!=\!\frac{\sqrt{2m(U_1-E)}}{2m},\: k_2\!=\!\frac{\sqrt{2m(E-U_2)}}{\hbar},\: \varkappa_3\!=\!\frac{\sqrt{2m(U_3-E)}}{\hbar}.
 \end{eqnarray}
 One also has to apply boundary and matching conditions:
 \begin{eqnarray}
 \psi_1(-\infty)=0,&&\qquad \psi_1(0)=\psi_2(0),\nonumber\\
 \psi_1^{'}(0)=\psi_2^{'}(0),&& \qquad \psi_2(0)=\psi_3(0),\nonumber\\
 \psi_2^{'}(0)=\psi_3^{'}(0),&&\qquad \psi_3(\infty)=0.
 \end{eqnarray}
 The first condition and the last one lead to $C_{12}=0$ and $C_{31}=0$.
 The rest conditions give
 \begin{eqnarray}
 \!\!\!\!\!&&C_{11}=C_{21},\qquad
 C_{11}\varkappa_1=C_{22}k_2,\nonumber\\
 \!\!\!\!\!&&C_{21}\cos(k_2a)+C_{22}\sin(k_2a)=C_{32}\exp(-\varkappa_3a),\nonumber\\
 \!\!\!\!\!&&-C_{21}k_2\sin(k_2a)+C_{22}k_2\cos(k_2a)=-\varkappa_3C_{32}\exp(-\varkappa_3a)
 \end{eqnarray}
 or
 \begin{eqnarray}
 k_2\frac{\varkappa_1-k_2\tan(k_2a)}
 {k_2+\varkappa_1\tan(k_2a)}=-\varkappa_3,
 \end{eqnarray}
 which finally gives the relation for a determination of energies of bound states
 \begin{eqnarray}\label{nsw_dr}
 \tan(k_2a)=\frac{k_2(\varkappa_1+\varkappa_3)}{\varkappa_1\varkappa_3-k_2^2}.
 \end{eqnarray}
 
 To determine the wave functions of bound states in this system we depict  it in  the following form
 \begin{eqnarray}
 \psi(x)=
 \left\{\begin{array}{cc}
 C_{11}\exp(\varkappa_1x), &x<0\\
 C_{21}\cos(k_2a)+C_{22}\sin(k_2a), &  0\leq x\leq 0\\
 C_{32}\exp(-\varkappa_3x), & x>a
 \end{array}\right..
 \end{eqnarray}
 To find constants $C_{11}$, $C_{21}$, $C_{22}$, $C_{32}$ we have to solve the following matrix equation:
 \begin{eqnarray}
 \left(\begin{array}{cccc}
 1 & -1 & 0 & 0  \\
 \varkappa_1 & 0 & -k_2 & 0  \\
 0 & \cos(k_2a)  & \sin(k_2a) &  -e^{-\varkappa_3a}\\
 0 & -k_2\sin(k_2a)  &  k_2\cos(k_2a) & \varkappa_3e^{-\varkappa_3a}
 \end{array}\right)
 \!\!\!
 \left(\!\begin{array}{c}
 C_{11}\\
 C_{21}\\
 C_{22}\\
 C_{32}
 \end{array}\!\right)\!\!=\!
 \left(\!\begin{array}{c}
 0\\
 0\\
 0\\
 0
 \end{array}\!\right)\!\!\!.
 \end{eqnarray}
 The existence of a non-trivial solution of this system of equations demands the determinant of this $4\times 4$ matrix to be equal to zero which bears
 the well-known relation (\ref{nsw_dr}). But it also means that values $C_{11}$, $C_{21}$, $C_{22}$, $C_{32}$ are not independent and three of them (for example $C_{21}$, $C_{22}$, $C_{32}$) can be expressed through the other one ($C_{11}$): 
 \begin{eqnarray}
 &&C_{21}=C_{11}, \qquad C_{22}=\frac{\varkappa_1}{k_2}C_{11}\nonumber\\
 &&C_{32}=C_{11}\exp(\varkappa_3a)\left\{\cos(k_2a)+\frac{\varkappa_1}{k_2}\sin(k_2a)\right\}.
 \end{eqnarray}  
 The final step is to determine $C_{11}$. One can do it using the normalization condition for a wave function of a bound state:
 \begin{eqnarray}
 |C_{11}|^2\left(\int\limits_{-\infty}^0\exp[2\varkappa_1x]dx+\int\limits_0^a
 \left[\cos(k_2x)+\frac{\varkappa_1}{k_2}\sin(k_2x)\right]^2dx+\right.\nonumber\\
 \left.+\exp(2\varkappa_3a)\left[\cos(k_2a)+\frac{\varkappa_1}{k_2}\sin(k_2a)\right]^2\int\limits_a^\infty \exp[-2\varkappa_3x]dx 
 \right)\!=1.
 \end{eqnarray}
 After an integration we get
 \begin{eqnarray}\label{C11_res_int}
 \!\!\!\!\!\!\!\!\!\!\!\!\!\!\!&&\frac{1}{2\varkappa_1}+\frac{1}{4k_2}\left(1-\frac{\varkappa_1^2}{k_2^2}\right)\sin[2k_2a]-\frac{\varkappa_1}{2k_2^2}\cos[2k_2a]+\nonumber\\
 \!\!\!\!\!\!\!\!\!\!\!\!\!\!\!&&+\!\frac{1}{2}\left(a\!+\!\frac{\varkappa_1^2a}{k_2^2}\!+\!\frac{\varkappa_1}{k_2^2}\right)\!+\!\frac{1}{2\varkappa_3}\left[\cos(k_2a)\!+\!\frac{\varkappa_1}{k_2}\sin(k_2a)\right]^2\!\!\!=\!\frac{1}{|C_{11}|^2}\!.
 \end{eqnarray}
 On the base of both a relation (\ref{nsw_dr}) and the following formulas:
 \begin{eqnarray}
 \sin(2k_2a)\!\!\!&=&\!\!\!\frac{2\tg(k_2a)}{\tg^2(k_2a)+1}=
 -\frac{2k_2(\varkappa_1\varkappa_3-k_2^2)(\varkappa_1+\varkappa_3)}{(\varkappa_1^2k_2^2+\varkappa_1^2\varkappa_3^2+k_2^4+k_2^2\varkappa_3^2)},\nonumber\\ 
 \cos(2k_2a)\!\!\!&=&\!\!\!\frac{1\!-\!\tg^2(k_2a)}{\tg^2(k_2a)\!+\!1}\!=\!
 -\frac{\varkappa_1^2k_2^2\!-\!\varkappa_1^2\varkappa_3^2\!+\!4\varkappa_1k_2^2\varkappa_3\!-\!k_2^4\!+\!k_2^2\varkappa_3^2}{\varkappa_1^2k_2^2+\varkappa_1^2\varkappa_3^2+k_2^4+k_2^2\varkappa_3^2}\nonumber\\
 \end{eqnarray} 
 one can transform the relation (\ref{C11_res_int}) to the following form 
 \begin{eqnarray}
 |C_{11}|^{-2}=\frac{1}{2}\left(1+\frac{\varkappa_1^2}{k_2^2}\right)
 \left(a+\frac{\varkappa_1+\varkappa_3}{\varkappa_1\varkappa_3}\right).
 \end{eqnarray}
 It gives us the final expression for the square of a modulus of a wave function in the well $(0<x<a)$:
 \begin{eqnarray}
 \!\!\!\!\!\!\!&&|\psi_2(x)|^2\!=\!2\left(1\!+\!\frac{\varkappa_1^2}{k_2^2}\right)^{-1}\!\!
 \left(a\!+\!\frac{\varkappa_1\!+\!\varkappa_3}{\varkappa_1\varkappa_3}\right)^{-1}\!\!\left(\cos(k_2x)\!+\!
 \frac{\varkappa_1}{k_2}\sin(k_2x)\right)^2\!\!=\nonumber\\
 \!\!\!\!\!\!\!&&=\left(a+\frac{\varkappa_1+\varkappa_3}{\varkappa_1\varkappa_3}\right)^{-1}\left(\frac{\cos(k_2a)}{\sqrt{1+\varkappa_1^2/k_2^2}}-\frac{-\varkappa_1/k_2
 	\sin(k_2a)}{\sqrt{1+\varkappa_1^2/k_2^2}}\right)^2=\nonumber\\
 \!\!\!\!\!\!\!&&=\left(a+\frac{\varkappa_1+\varkappa_3}{\varkappa_1\varkappa_3}\right)^{-1}\cos^2(k_2x+\phi),
 \end{eqnarray}
 where 
 \begin{eqnarray}
 \cos(\phi)=\frac{1}{\sqrt{1+\varkappa_1^2/k_2^2}},\qquad 
 \sin(\phi)=-\frac{\varkappa_1/k_2}{\sqrt{1+\varkappa_1^2/k_2^2}}.
 \end{eqnarray}
   
 \section{Transfer matrix approach}
 To apply the transfer matrix approach we have to built three matrices and then to multiply them. The first matrix $I_{12}$ describes a wave transition through the interface ($x=0$) between the first and the second region of a potential (\ref{UU1U2U3}):
 \begin{eqnarray}
 I_{12}=\frac{1}{2} 
 \begin{pmatrix}
 \frac{\varkappa_1+ik_2}{ik_2}, & \frac{ik_2-\varkappa_1}{ik_2} 
 \\
 \frac{ik_2-\varkappa_1}{ik_2}, & \frac{\varkappa_1+ik_2}{ik_2}
 \end{pmatrix}.
 \end{eqnarray}
 A wave transferring inside the second region of a potential (\ref{UU1U2U3}) is described by a second matrix, namely matrix $M$
 \begin{eqnarray}
 M=\begin{pmatrix}
 e^{ik_2a}, & 0 
 \\
 0, & e^{-ik_2a}
 \end{pmatrix}.
 \end{eqnarray}
 And finally the transition through the interface ($x=a$) between a second and a third region of a potential (\ref{UU1U2U3}) is described by a matrix $I_{23}$:
 \begin{eqnarray}
 I_{23}=\frac{1}{2}\begin{pmatrix}
 \frac{ik_2+\varkappa_3}{\varkappa_3} & \frac{\varkappa_3-ik_2}{\varkappa_3} 
 \\
 \frac{\varkappa_3-ik_2}{\varkappa_3} & \frac{ik_2+\varkappa_3}{\varkappa_3}
 \end{pmatrix}.
 \end{eqnarray}
 Thus, the full transfer matrix has the following form
 \begin{eqnarray}
 \!\!\!\!\!\!\!\!\!&&T\!=\!I_{12}MI_{23}\!=\!\frac{1}{4} 
 \!\begin{pmatrix}
 \frac{\varkappa_1+ik_2}{ik_2} &\!\!\! \frac{ik_2-\varkappa_1}{ik_2} 
 \\
 \frac{ik_2-\varkappa_1}{ik_2} &\!\!\! \frac{\varkappa_1+ik_2}{ik_2}
 \end{pmatrix}
 \!\!
 \begin{pmatrix}
 e^{ik_2a} & \!\!\!0 
 \\
 0 & \!\!\!e^{-ik_2a}
 \end{pmatrix}
 \!\!
 \begin{pmatrix}
 \frac{ik_2+\varkappa_3}{\varkappa_3} & \!\!\!\frac{\varkappa_3-ik_2}{\varkappa_3} 
 \\
 \frac{\varkappa_3-ik_2}{\varkappa_3} & \!\!\!\frac{ik_2+\varkappa_3}{\varkappa_3}
 \end{pmatrix}\!\!=\nonumber\\
 \!\!\!\!\!\!\!\!\!&&=\!\!\begin{pmatrix}
 \frac{k_2(\varkappa_1\!+\!\varkappa_3)\!\cos(k_2a)\!+\!(\varkappa_1\varkappa_3\!-\!k_2^2)\!\sin(k_2a)}{2k_2\varkappa_3} &\!\!\! \frac{k_2(\varkappa_3\!-\!\varkappa_1)\!\cos(k_2a)\!-\!(\varkappa_1\varkappa_3\!+\!k_2^2)\!\sin(k_2a)}{2k_2\varkappa_3}
 \\
 \frac{k_2(\varkappa_3\!-\!\varkappa_1)\!\cos(k_2a)\!+\!(\varkappa_1\varkappa_3\!+\!k_2^2)\!\sin(k_2a)}{2k_2\varkappa_3}
 &\!\!\! \frac{k_2(\varkappa_1\!+\!\varkappa_3)\!\cos(k_2a)\!-\!(\varkappa_1\varkappa_3\!-\!k_2^2)\!\sin(k_2a)}{2k_2\varkappa_3}
 \end{pmatrix}\!.\nonumber\\
 \!\!\!\!\!\!\!\!\!
 \end{eqnarray}
 The relation for a determination of energies of bound states is $M_{11}=0$ or in the explicit form
 \begin{eqnarray}
 \frac{k_2(\varkappa_1+\varkappa_3)\cos(k_2a)+(\varkappa_1\varkappa_3-k_2^2)\sin(k_2a)}{2k_2\varkappa_3}=0,
 \end{eqnarray}
 which gives the same expression as in the previous section
 \begin{eqnarray}
 \tan(k_2a)=\frac{k_2(\varkappa_1+\varkappa_3)}{\varkappa_1\varkappa_3-k_2^2}.
 \end{eqnarray}

\section{Conclusions}
 Making comparision of three different approaches which were described in this paper we can conclude that a 
 quantum wave impedance approach for a calculation of energies of bound states and their  wave functions demands much less efforts than a classical method or a transfer matrix technique. Thus  a quantum wave impedance app\-roach is a powerfull method for studying quantum mechanical systems  and, in particular, can be applied for nanosystems with a complicated geomry of a potential energy. But it is also an elegant teaching and learning tool since it has  quite  simple both a physical interpretation and a mathematical technique.

It is worth to say that the paper \cite{Nelin_Vodolazka:2014} is dedicated to the similar problem, namely to the comparative analysis of traditional and impedance approaches in a modelling
asymmetric potential quantum-mechanical barrier. As a result the authors stated that ``impedance method significantly simplifies modelling of quantum-mechanical
struc\-tures in comparison with the traditional method of solving quantum-mechanical problems''. But that paper does not consider nor a transfer matrix approach nor finding wave functions of bound states. This task we realized in this paper with hope that it will demonstrate other sides of a quantum wave impedance approach and its advantages.

\renewcommand\baselinestretch{1.0}\selectfont


\def\name{\vspace*{-0cm}\LARGE 
	Bibliography\thispagestyle{empty}}
\addcontentsline{toc}{chapter}{Bibliography}

{\small

	\bibliographystyle{gost780u}
	\bibliography{full.bib}
	
}

\newpage

\end{document}